\documentclass[12pt,preprint]{aastex}

\begin{document}
\input colordvi

\title{X-ray Emission of Baryonic Gas in the Universe:
   Luminosity-Temperature Relationship and Soft-Band Background}

\author{Tong-Jie Zhang$^{1,2}$, Jiren Liu$^{3}$,
Long-long Feng$^{3,5}$, Ping He$^{4}$, Li-Zhi Fang$^{2}$}

\altaffiltext{1}{Department of Astronomy, Beijing Normal
University, Beijing 100875, China; tjzhang@bnu.edu.cn}

\altaffiltext{2}{Department of Physics, University of Arizona,
Tucson, AZ 85721; fanglz@physics.arizona.edu, tzhang@physics.arizona.edu}

\altaffiltext{3}{Purple Mountain Observatory, Nanjing 210008,
China; liujr@mail.ustc.edu.cn, fengll@pmo.ac.cn}

\altaffiltext{4}{Institute of Theoretical Physics, Chinese Academy
of Science, Beijing, P.R.China; hep@itp.ac.cn}

\altaffiltext{5}{National Astronomical Observatories, Chinese
Academy of Science, Chao-Yang District, Beijing 100012,
China}

\begin{abstract}

We study the X-ray emission of baryon fluid in the universe using
the WIGEON cosmological hydrodynamic simulations. It has been
revealed that cosmic baryon fluid in the nonlinear regime behaves
like Burgers turbulence, i.e. the fluid field consists of shocks.
Like turbulence in incompressible fluid, the Burgers turbulence
plays an important role in converting the kinetic energy of the fluid
to thermal energy and heats the gas. We show that the simulation
sample of the $\Lambda$CDM model without adding extra heating sources can
fit well the observed distributions of X-ray luminosity versus
temperature ($L_{\rm x}$ vs. $T$) of galaxy groups and is also consistent with
the distributions of X-ray luminosity versus velocity dispersion
($L_{\rm x}$ vs. $\sigma$).
Because the baryonic gas is multiphase, the $L_{\rm x}-T$ and
$L_{\rm x}-\sigma$ distributions are significantly scattered. If
we describe the relationships by power laws $L_{\rm x}\propto
T^{\alpha_{LT}}$ and $L_{\rm x}\propto \sigma^{\alpha_{LV}}$, we
find $\alpha_{LT}>2.5$ and $\alpha_{LV}>2.1$. The X-ray background
in the soft $0.5-2$ keV band emitted by the baryonic gas in the
temperature range $10^5<T<10^7$ K has also been calculated. We
show that of the total background, (1) no more than 2\% comes
from the region with temperature less than $10^{6.5}$ K, and
(2) no more than 7\% is from the region of dark matter with mass
density $\rho_{\rm dm}<50 \bar{\rho}_{\rm dm}$. The region of
$\rho_{\rm dm}>50\bar{\rho}_{\rm dm}$ is generally clustered and
discretely distributed. Therefore, almost all of the soft X-ray
background comes from clustered sources, and the contribution from
truly diffuse gas is probably negligible. This point agrees with
current X-ray observations.

\end{abstract}


\section{Introduction}

In the linear regime of gravitational clustering of cosmic matter,
the evolution of baryonic gas follows dark matter point by point
(Bi et al. 1992; Fang et al. 1993; Nusser \& Haehnelt 1999; Nusser
2000). That is, the density and velocity distributions of baryonic
matter can be obtained by a similar mapping from the dark matter
field. The similar mapping has also been used in modeling the gas
in clusters of galaxies (Kaiser 1986). However, it has been found
that the similarity between the mass and velocity fields of
baryonic gas and dark matter is broken in clusters that are
formed through highly non-linear evolution. The mechanism of the
breaking of similarity is due to the different dynamical behavior of
dark matter and baryonic gas. The former is collisionless, while
the latter is approximately an ideal fluid. The velocity field of
collisionless dark matter particles is multivalued at the
intersection of their trajectories, while baryon fluid always has
a single-value velocity field. Thus, shocks in baryon fluid will
appear at the intersection and break the similarity between the
mass and velocity fields of baryon fluid and dark matter
(Shandarin \& Zel'dovich 1989).

At later times, it has been recognized that shocks occur not only
in high-, but also in middle- and even low-density regions. This
point has been shown by the dynamical equation of baryonic gas.
Although cosmic baryonic gas is a Navier-Stokes fluid, its
evolution is dominated by the growth mode, which is
approximately governed by a random-force-driven Burgers equation,
and the random force is produced by the gravity of the random
field of dark matter (Gurbatov et al 1989; Berera \& Fang 1994;
Vergassola et al. 1994; Jones 1999; Matarrese and Mohayaee 2002;
Pando et al. 2002, Pando et al 2004). When the Reynolds number is
large, shock-caused turbulence, called Burgers turbulence, will
develop in the fluid (L\"assig 2000). The dynamics and
thermodynamics of baryonic gas will be substantially affected by
the Burgers turbulence. Dynamically, it will lead to the
discrepancy between baryonic matter and dark matter, like the
discrepancy of a passive substance from the underlying field
during nonlinear evolution (e.g. Shraiman \& Siggia 2000).
Thermodynamically, the Burgers turbulence leads to the
multiple phases of thermal properties and to converting the bulk
kinetic energy of the fluid to thermal energy. For the system of
cosmic baryonic and dark matter, a large Reynolds number actually
corresponds to the onset of the non-linear regime of the
gravitational clustering. Therefore, the similarity between baryonic
and dark matter will inevitably break in the non-linear regime,
and the dynamical and thermodynamical effects of the Burgers
turbulence must be considered.

Some effects of the Burgers turbulence have been detected with
observational data and/or hydrodynamic simulation samples of
cosmic baryonic gas. First, the statistical decoupling between the
density and velocity fields of baryonic gas and dark matter is
found to be significant on scales larger than the Jeans length
(Pando et al 2004; Kim et al 2005). The importance of the
discrepancy in the early nonlinear evolution has also been noted
by Yoshida et al. (2003). Second, the transmitted flux of QSOs'
Ly$\alpha$ absorption is found to be remarkably intermittent
(Jamkhedkar et al 2000, 2003, 2005; Pando et al 2002; Feng et al
2003), which is consistent with the prediction of the
intermittence of fully developed Burgers turbulence (Polyakov
1995; Balkovsky et al. 1997; Frisch et al 2001). Third, the
temperature field of the baryonic gas is multiphase, and the
heating caused by the shocks of the Burgers turbulence is
substantial. Consequently, the distribution of the baryon fraction on
large scales is nonuniform (He et al. 2005), and high-entropy gas
is produced (He et al. 2004).

In this paper, we extend these studies to the X-ray emission of the
hot baryonic gas in the universe. X-ray emission of galaxy groups
is considered as important evidence of the similarity
breaking between baryon fluid and dark matter. The relation
between X-ray luminosity $L_{\rm x}$ and temperature $T$ predicted
by the similarity takes the form of a power law, $L_{\rm x}\propto
T^{\alpha}$, with an index of $\alpha=2$ (Kaiser 1991), while the
observed result is $\alpha > 2$ (Edge \& Steward 1991; David et
al. 1993; Wu et al 1999; Helsdon \& Ponman 2000; Xue \& Wu 2000;
Croston et al. 2005). Moreover, the X-ray background in the soft band
given by similarity is found to be much higher than the observed
upper limit. In order to solve these problems, various models of
non-gravitational heating for baryonic gas have been proposed (e.g.
Valageas \& Silk 1999; Tozzi \& Norman 2001; Voit et al 2002;
Zhang \& Pen 2003; Xue \& Wu 2003). The common goal of these
models is to violate the similarity by heating baryonic gas before
it falls into gravity potential wells. The amount of heating is
of the order of 1 keV nucleon$^{-1}$ (Pen 1999; Wu et al 1999).

However, most non-gravitational heating models {\it assume}
that the similarity will still hold without the non-gravitational
heating sources. This point, as pointed out above, is actually
inconsistent with the dynamics of the cosmic baryonic gas in the
non-linear regime. Baryonic gas, either in high- or in low-density
regions, will be heated when Burgers turbulence develops,
regardless of whether non-gravitational heatingi is added. Typical shocks of
the Burgers turbulence can convert the kinetic energy of a baryon
fluid with a speed of a few hundred km s$^{-1}$ into thermal
energy (He et al. 2004; Kim et al. 2005), which is of the order of
1 keV nucleon$^{-1}$. Therefore, the thermodynamical effect of the
Burgers turbulence should be substantial on the $L_{\rm x}-T$
relation and soft X-ray background. We need, at least, to estimate
the effect of the hydrodynamic heating on the X-ray emission
before considering extra non-gravitational heating sources.

The paper is organized as follows. \S 2 describes the features of
the baryonic gas as a Burgers fluid with samples of hydrodynamic
cosmological simulation. \S 3 presents the analysis of the
relationships between X-ray luminosity and mass density,
temperature, and velocity dispersion. The soft X-ray background
radiation is addressed in \S 4. Finally, the conclusions and
discussion are given in \S 5.

\section{Baryonic gas and Burgers turbulence}

\subsection{Gravitational clustering and Burgers fluid}

The baryonic gas is generally assumed to be an ideal fluid
satisfying the hydrodynamic equations
\begin{equation}
\frac{\partial \delta}{\partial t} +
  \frac{1}{a}\nabla \cdot (1+\delta) {\bf v}=0
\end{equation}
\begin{equation}
\frac{\partial a{\bf v}}{\partial t}+
 ({\bf v}\cdot \nabla){\bf v}=
-\frac{1}{\rho}\nabla p - \nabla \phi
\end{equation}
\begin{equation}
\frac{\partial {\cal E}}{\partial t}+5\frac{\dot{a}}{a}{\cal E}+
  \frac{1}{a}\nabla\cdot ({\cal E}{\bf v})=
   -\frac{1}{a}\nabla\cdot(p{\bf v})-
   \frac{1}{a}\rho_{\rm igm}{\bf v}\cdot \nabla\phi- \Lambda_{\rm rad},
\end{equation}
where $\rho$, ${\bf v}$, ${\cal E}$, and $p$ are
the mass density, peculiar velocity, energy density, and pressure
of the gas respectively. The cosmic factor $a$ describes the cosmic
expansion.
The term $\Lambda_{\rm rad}$ in Eq.(3) is given by
radiative heating and cooling of the baryonic gas per unit volume.
The gravitational potential $\phi$ in Eqs.(2) and (3) is given by
\begin{equation}
\nabla^2 \phi = 4\pi G a^2\bar{\rho}_{\rm dm}\delta_{\rm dm},
\end{equation}
where $\bar{\rho}_{\rm dm}(t)$ and $\delta_{\rm dm}$ are, respectively,
the mean mass density and density contrast of the perturbations of
dark matter. Here we assume that the gravitational potential $\phi$
is only produced by the dark matter mass perturbation.

During the process of gravitational clustering, there are two types
of modes for the perturbations of the peculiar velocity field. For
growth mode, the velocity field is curl-free, and the vertical mode
is generally decaying with time. Therefore, one can define a
velocity potential  $\varphi$ by ${\bf\it v}=-(1/a)\nabla \varphi$.
To sketch the gravitational clustering, we consider the case that
all the thermal processes are approximated by the polytropic
relations $p \propto \rho^{\gamma}$, $T \propto \rho^{\gamma-1}$,
or $T =T_0(1+\delta)^{\gamma-1}$, where $\delta({\bf x},t)
=[\rho({\bf x},t)-\bar{\rho}(t)]/\bar{\rho}(t)$ is the baryon mass
density perturbation. In this case, the equation of velocity
potential, Eq.(3), can be approximately rewritten as
\begin{equation}
\frac{\partial \varphi}{\partial t}-
\frac{1}{2a^2}(\nabla \varphi)^2 -
\frac{\nu}{a^2}\nabla^2 \varphi =\phi,
\end{equation}
where the coefficient $\nu/a^2=1/k^2_{\rm J}$ acts like a viscosity
due to thermal diffusion and is characterized by the Jeans length
$k_{\rm J}^2=(a^2/t^2)(\mu m_{\rm p}/\gamma k_{\rm B}T_{\rm 0})$.

Eq.(5) is a stochastic-force-driven Burgers equation, which
contains two scales: the dissipation length or the Jeans length
$1/k_{\rm J}$ and the correlation length of the gravity potential
$\phi$, $r_{\rm c}$. The intensity of $\phi$ can be quantified by
the density contrast of dark matter $\delta_{\rm dm}$. A basic
feature of the Burgers equation is that turbulence will develop in
the fluid if the following condition holds (e.g.\, Feng et al
2003):
\begin{equation}
(k_{\rm J}r_{\rm c})^{2/3}\langle \delta_{\rm dm}^2\rangle^{1/3}>1,
\end{equation}
which corresponds to the condition that the Reynolds
number must be larger than 1 for the turbulence of
incompressible fluid. Therefore, the quantity on the left-hand
side of Eq.(6) plays the same role as the Reynolds number in the
original fluid (L\"assig 2000).

Burgers turbulence is qualitatively different from the turbulence of
incompressible fluid. The latter generally consists of vortices on
various scales, while the former is a collection of shocks. These
features arise because, for growth modes, the fluid is potential
and the velocity field is irrotational. If $\nu \rightarrow 0$,
the velocity field in the Burgers prescription acquires
singularities due to the discontinuities caused by strong shocks.
The nonlinear feature of the velocity field of baryonic gas can be
understood as a field consisting of these shocks.

\subsection{Samples of hydrodynamic simulation}

In the numerical calculation, we still use the hydrodynamic
equations (1)-(3) to model the baryonic gas, because the decaying
modes will disappear automatically, and the result is just on the
growth modes. Since the field of baryonic gas in the nonlinear
regime consists of strong and weak shocks, the hydrodynamic
cosmological simulation should be able to capture shocks and
calculate the thermal properties of the baryon fluid in front of
and behind shocks. Therefore, an optimal simulation scheme has to
satisfy two conditions: (1) it is effective in capturing shock and
discontinuity transitions, and (2) it accurately calculates
piecewise smooth functions with a high resolution. Condition
(1) is obvious in studying a field consisting of shocks. 
Condition (2) is important in calculating the shock heating, which
depends on the difference between gaseous dynamical and thermal
properties of pre- and postshock regions.

We take the WIGEON (Weno for Intergalactic medium and Galaxy
Evolution and formatiON) code (Feng et al. 2004), which is a
cosmological hydrodynamic/$N$-body code based on the Weighted
Essentially Non-Oscillatory (WENO) algorithm (Harten et al. 1986;
Liu et al. 1994; Jiang \& Shu 1996; Shu 1998; Fedkiw, Sapiro \&
Shu 2003; Shu 2003). The WENO algorithm is an Eulerian approach. It has
been applied to hydrodynamic problems containing strong shocks,
complex structures, and turbulence. It has also been applied to the 
Burgers equation (Shu 1999). The WIGEON code has passed the necessary
tests including the Sedov blast wave and the formation of the
Zel'dovich pancake (Feng et al. 2004). It is successful in showing the
features of the Burgers fluid with the velocity field of cosmic
baryonic gas. That is, when the Reynolds number is high, the
velocity field consists of an ensemble of shocks with the
following features: (1) the probability distribution function
(PDF) of velocity is asymmetric between acceleration and
deceleration events and (2) the PDF of velocity difference $\Delta
v=v(x+r)-v(x)$ satisfies the scaling relation for a Burgers fluid
(Kim et al. 2005; He et al. 2005).

In order to accurately calculate the shock heating, the resolution
of the simulation should be less than the thickness of the shock,
which is of the order of the Jeans diffusion, which is greater
than $0.2$ $h^{-1}$ Mpc for redshifts $z<2$ (Bi et al. 2003). On
scales less than the Jeans diffusion, shock-caused discontinuity
is small, and the heating effect is weak. We perform three sets of
simulations: two (samples A and B) are in a periodic cubic box
with a size of 100 $h^{-1}$ Mpc, a 512$^3$ grid and an equal number of
dark matter particles, and another one (sample C) is in a cubic
box with a size of 25 $h^{-1}$ Mpc, a 192$^3$ grid and an equal
number of dark matter particles. The sizes of the grids are
$100/512=0.20$ and $25/192=0.13$ $h^{-1}$ Mpc, respectively. 
Sample C is the same as that used in the analysis of the Burgers
fluid features of cosmic baryonic gas (Pando et al 2004; Kim et al
2005).

We use the standard $\Lambda$CDM model, which is specified by the matter
density parameter $\Omega_{\rm m}=0.27$, baryonic matter density
parameter $\Omega_{\rm b}=0.044$, cosmological constant
$\Omega_{\Lambda}=0.73$, Hubble constant $h=0.71$, and mass
fluctuation $\sigma_8=0.84$ within a sphere of radius 8 $h^{-1}$
Mpc. The transfer function is calculated using CMBFAST (Seljak \&
Zaldarriaga 1996). For sample C, we use the same parameters
but $\Omega_{\rm b}=0.026$. We use the cloud-in-cell method for
mass assignment and interpolation and adopt the seven-point finite
difference to approximate the Laplacian operator. The simulations
start at the redshift $z=49$, and the results output at redshifts
$z$=2, 1, 0.5, and 0.

For the two simulations of the 100 $h^{-1}$ Mpc cubic box, one
considered the metal abundance (sample A), and one used the primordial
composition (sample B). For sample C, we used the primordial
composition, for which the atomic process of H and He
($X$=0.76,$Y$=0.24) is calculated as in Theuns et al. (1998). For
sample A, the metal cooling and metal line emission is calculated
by the phenomenological method (1) assuming a uniform evolving
metallicity $Z=0.3 Z_{\sun}(t/t_0)$,  where $t_0$ is the present
universe age, and (2) computing the cooling function using the table of
Sutherland \& Dopita (1993). An evolving background UV spectrum
calculated by Haardt \& Madau (1996) is used. With the comparison
among samples A, B, and C, one can estimate the effects of the size
of the simulation box and the metal abundance.

A common problem of hydrodynamic simulation with Eulerian variables 
is that it cannot describe compact objects on scales less than
the size of the grid. Therefore, we do not use these samples to
study rich clusters. On the other hand, the samples with Eulerian
variables make it easy to reach low-density regions. They are suitable
for analyzing X-ray emission of baryonic gas from background and
weakly clustered structures, such as groups.

\subsection{Shock heating and similarity breaking}

Shock heating is well known in the heating problem of groups,
clusters (Cavaliere et al 1997), and large-scale structures
(Miniati et al 2004; Kang et al. 2004). The scenario of Burgers
turbulence shows that, in nonlinear regime, shocks or complex
structures of the velocity field form around massive halos, as well as
regions with moderate and even low mass density. Therefore, it
heats the gas in various density fields. Figure 1 presents
the relations between the temperature and mass density of baryonic
gas at redshifts $z=$2, 1, 0.5, and 0.0. Each panel contains
19,200 data points randomly drawn from sample C. Figure 1 shows
that the gas is multiphase. It contains the phase of tightly
correlated $T$ and $\rho$ in the region of $T \leq10^{4.5}$K and
$0.01<\rho_{\rm igm}<2$ and also a phase of the scattered
distribution of $T$ ($>10^{5}$K) with respect to a given $\rho$.
Comparing with other numerical simulation results (e.g., Ryu et
al. 1993; Katz, Weinberg, \& Hernquist 1996; Theuns et al. 1998;
Dav\'e et al. 1999; Valageas, Schaeffer, Silk 2002; Springel \&
Hernquist 2002), the shock-heated gas is more significant in the
moderate- and low-density ($\rho_{\rm igm}\simeq 1$) regions. This
is because the WENO code is effective in capturing shocks in both
high- and low-density regions.

A dynamical effect of the Burgers turbulence is the discrepancy
between the fields of baryonic gas and dark matter. Figure 2 shows
two-dimensional contours of the baryonic gas density $\rho_{\rm
igm}$ and dark matter density $\rho_{\rm dm}$, both of which are in
units of their corresponding average density $\bar\rho_{\rm igm}$
and $\bar\rho_{\rm dm}$. One can clearly see from Figure 2 that
the density of baryonic gas ({\it right}) is not simply
proportional to that of dark matter ({\it left}). The size of the 
over-density region $\rho_{\rm igm}>1$ is generally larger than
the dark matter counterpart. Therefore, the clumping of baryonic
gas is generally weaker than that of dark matter.

Figure 3 shows the same slice as in Figure 2, but gives the
two-dimensional contours of temperature and the ratio between the
densities of baryonic matter and dark matter. The left panel shows
that the contours of $\rho_{\rm igm}/\rho_{\rm dm}>1$ are located
outside the massive halos. Contrarily, the center of the halos has
only a low baryon fraction $\rho_{\rm igm}/\rho_{\rm dm}<1$. It is
more interesting to see from the right panel that the shape of
high-temperature regions is very different from high-density
regions of either dark matter or baryonic matter. In Figure 3, the
high-density regions of $\rho_{\rm igm}$ or $\rho_{\rm dm}$
consist of filaments and knots, while the high-temperature
regions are isolated spots. Therefore, temperature is not simply
proportional to $\rho_{\rm igm}$ or $\rho_{\rm dm}$, but
multiphase.

In summary, with respect to the similarity model, the non-linear
evolution of cosmic baryon fluid leads to (1) a shallower
distribution of baryonic gas, (2) a lower density $\rho_{\rm igm}$
inside the dark matter halo, and (3) a higher temperature outside dark
matter halos. Generally, various non-gravitational heating models
try to realize the above-mentioned features. Figures 1-3 show,
however, that all of these features are naturally yielded when
baryonic gas is undergoing a Burgers turbulence evolution.

\section{X-ray emission of baryonic gas from groups of galaxies}

\subsection{Relationship between X-ray luminosity and mass density}

For the baryonic gas in each cell of the grid with temperature
$T>10^5$ K, we can calculate the X-ray luminosity by
the bremsstrahlung process assuming that all H and He atoms are
ionized, and then obtain a field of luminosity $L_{\rm x}$. In
Figure 4, we plot, respectively, the relationships between the
X-ray luminosity and mass density of baryonic gas at redshifts $z$=0.0, 0.5,
1.0, and 2, which are given by 9931, 9119, 7314, and 4864 data
points randomly selected from sample C. Here we consider only
the bremsstrahlung radiative process, without considering the
effect of metal line emission, which may lead to an uncertainty of
no more than about 20$\%$.

\begin{deluxetable}{cccccc}
\tablecolumns{8}
\tablewidth{0pc}
\tablecaption{Mean intensity of X-ray luminosity ($10^6<T<10^7$K)}
\tablehead{
 redshift & total & $\rho_{\rm dm}<100$ & $\rho_{\rm dm}<50$ &
$\rho_{\rm dm}<10$\cr & \multicolumn{4}{c}{(in unit of $10^{43}$
ergs s$^{-1}$)}
}
\startdata
 $z=2.0$ & 9.10$\times 10^{-8}$ & 4.12$\times 10^{-8}$
    & 2.69$\times 10^{-8}$ & 5.39$\times 10^{-9}$ \\
$z=1.0$   & 1.12$\times 10^{-7}$ & 4.41$\times 10^{-8}$ &
       2.77$\times 10^{-8}$ & 5.04$\times 10^{-9}$ \\
$z=0.0$   & 2.74$\times 10^{-7}$ & 5.27$\times 10^{-8}$ &
       3.28$\times 10^{-8}$ & 5.63$\times 10^{-9}$\\
\enddata
\end{deluxetable}

Figure 4 shows that for $z<2$, the $L_{\rm x}-\rho_{\rm
igm}$ distribution is weakly dependent on redshift, and X-ray
luminosity is correlated with the density $\rho_{\rm igm}$.
Considering the adiabatic ``equation of state" $T\propto
\rho^{2/3}$ (e.g. He et al 2004), we then obtain $L_{\rm x}
\propto \rho_{\rm igm}^2T^{1/2} \propto \rho^{2.3}$. However, the
$L_{\rm x}$-$\rho_{\rm igm}$ distribution is actually scattered
due to the multiphase property of baryonic gas. For a given $\rho_{\rm
igm}$, the scatter of baryonic gas temperature is of the order of
$10^3$, i.e. from $10^5$ to $10^7$ K. Therefore, the scatter of
$L_{\rm x}$ for a given $\rho_{\rm igm}$ can be as large as a
factor of 10$^{1.5}$.

Table 1 gives the mean X-ray luminosity for sample C from
regions with temperature $10^6<T<10^7$K and dark matter density
$\rho_{\rm dm} <100$, 50, and 10. It shows that the X-ray luminosity 
is mostly produced from the epoch of $z\leq 1$. At $z=1$, about
25\% of the X-ray emission is from moderate and low clustering regions
$\rho_{\rm dm}<50$, while at $z=0$, only about 10\% is
from moderate- and low-clustering areas.

Figure 5 presents the $L_{\rm x}-\rho_{\rm dm}$ relationship for
the $z=0$ sample, but the fields of $\rho_{\rm dm}$ and $T$ are
decomposed to the cells on scales of 0.52, 1.04, and 2.08 $h^{-1}$
Mpc. We use the scaling function of the discrete
wavelet transform (DWT) to perform the decomposition. The DWT
decomposition does not cause false correlation (Fang \& Thews
1998). Therefore, $L_{\rm x}$ of Figure 5 is the total luminosity
from the cells on scales of 0.52, 1.04, and 2.08 $h^{-1}$ Mpc, and
$\rho_{\rm dm}$ is the mean density of dark matter in these cells.

The cells on scales of 0.52, 1.04, and 2.08 $h^{-1}$ Mpc with density
$\rho_{\rm dm}>50$ have total masses $\geq 0.27, 2.18$, and $17.4
\times 10^{12}$ M$_{\odot}$, respectively. These masses and sizes
correspond to groups or clusters. It has been shown that the
DWT-identified cells on scale of 1.5 $h^{-1}$ Mpc with high density
are statistically the same as clusters identified by a traditional
method, such as the friends-of-friends algorithm (Xu et al. 1998).
Moreover, the cells with high density are found to be virialized
or quasi-virialized (Xu et al 2000). Therefore, the DWT-identified
cells with $\rho_{\rm dm}\geq 50$ on a scale of 1.04 h$^{-1}$ Mpc or
$\rho_{\rm dm}\geq 100$ on a scale of 0.52 h$^{-1}$ Mpc give an ensemble of
groups and clusters. This can be applied to estimate the statistical
properties of galaxy groups. The cells with high $L_{\rm x}$
simulate the behavior of the X-ray emission of galaxy groups.

Generally, the mean of the luminosity $L_{\rm x}$ of Figure 5
becomes larger with the increase of the cell scale. This is
trivial, because the volume of a 1.04 (2.08) $h^{-1}$ Mpc cell is 8
times larger than that of a 0.52 (1.04) $h^{-1}$ Mpc cell, and so on. An
interesting feature shown in Figure 5 is that the scatter of the
$L_{\rm x}-T$ distributions does not reduce with the increase of
cell scale, even when the smoothing scale of 2.08 $h^{-1}$ Mpc is
larger than the largest size of virialized collapsed objects.
Therefore, the scatter of luminosity should be not only from
collapse, but dependent on the multiphase property of the temperature
field of baryonic gas.

\subsection{Relationship between X-ray luminosity and temperature}

Figure 6 presents the distribution of $L_{\rm x}$ versus $T$ for the
simulation sample A at $z=0$ on a scale of 0.78 $h^{-1}$ Mpc. The
observed $L_{\rm x}-T$ distribution of galaxy groups is also shown
in Figure 6. The data are taken from Helsdon \& Ponman (2000), Xue
\& Wu (2000), and Croston et al (2005). The three observed samples
are partially overlapped. If a group is listed in two or three
samples, we only use the data of the latest one.

Figure 6 shows that the simulated results are in good agreement
with the observed $L_{\rm x}-T$ distribution of groups. First, the
simulated points of high $L_{\rm x}$ and $T$ are located in the
same areas as in the observed $L_{\rm x}-T$ distribution. As mentioned 
in \S 3.1 above, the high $L_{\rm x}$ cells are virialized or
quasi-virialized objects with masses of groups. Second, the observed
$L_{\rm x}-T$ distribution is connected smoothly and continuously
with simulation points with lower $L_{\rm x}$ and $T$. The cells
with lower $L_{\rm x}$ have lower density $\rho_{\rm igm}$ or
$\rho_{\rm dm}$. Therefore, Figure 6 strongly indicates that the
X-ray emission from baryonic gas related to groups underwent the
same dynamical and thermodynamical evolution as that of the diffuse
clouds, which have weaker X-ray emission than groups.

Similar to the distribution of $L_{\rm x}$ versus $\rho_{\rm igm}$ and
$\rho_{\rm dm}$ (Figures 4 and 5), the $L_{\rm x}-T$ distribution 
is largely scattered.  This
scatter cannot be explained by the Jeans diffusion or another 
Gaussian noise process. Figure 6 shows an upper envelope in the
$L_{\rm x}-T$ distributions, which corresponds to the maximum gas
density $\rho_{\rm igm}$ for a given temperature $T$. The upper
envelope can be approximately fitted by a power law $L_{\rm
x}\propto T^{2.5}$. Thus, we can conclude that if we try to fit
the $L_{\rm x}-T$ relationship of groups by a power law $L_{\rm
x}\propto T^{\alpha_{LT}}$, the index $\alpha_{LT}$ should be
$\geq 2.5$. This result is consistent with the measured
$\alpha_{LT}$ for samples of groups, which are $3.6
<\alpha<8.2$ (Helsdon \&  Ponman 2000), $2.1 <\alpha<5.7$ (Xue \&
Wu 2000), and $2.7<\alpha<4.1$ (Croston et al. 2005).

Figure 7 is the same as Figure 6, but for sample C. The DWT
variables are on scales of 0.52 and 1.04 $h^{-1}$ Mpc.
The simulation data in Figure 7 are also basically consistent with
the observed result. Therefore, the effect of metal cooling on the
$L_{\rm x}-T$ relation is small. As we know, the major effect of
the metal cooling is to form a condensation branch in the
temperature-density distribution. The temperature of the
condensation branch is low, and therefore, these are not strong
X-ray sources. It is interesting to see that Figure 7 shows a
lower envelope with a power law $L_{\rm x}\propto T^{\sim 5}$. This power
law is about the same as $L_{\rm x}\propto T^{\sim 4.8}$ given by the
hydrodynamic simulation of Dav\'e et al (2002). However, our
result of the $L_{\rm x}-T$ distribution is scattered, i.e. fills
in the range between the lower and upper envelopes. Observed
samples also show the scatter and are consistent with our
simulation sample.

Figure 8 presents the $L_{\rm x}-T$ distribution for sample C,
but decomposed on physical scales of 0.69 $h^{-1}$ Mpc at redshift
$z=0.5$ and 0.5 $h^{-1}$ Mpc at $z=1$. The observed
points are the same as those in Figures 6 and 7. We see that the
simulated $L_{\rm x}-T$ distribution is almost $z$ independent in
the range $z\leq 1$. This is consistent with a weak evolution
scenario of temperature and entropy fields in the range $z\leq 2$
(He et al. 2004). Thus, we may predict that the $L_{\rm x}-T$
distribution of groups at $z\sim 1$ should not be very different
from $z=0$.

\subsection{Relationship between X-ray luminosity
  and velocity dispersion}

The velocity dispersion in a cell can be estimated by the peculiar
velocity difference within this cell. We use the DWT variables to
calculate the velocity dispersion. The details of this algorithm
are given by Yang et al (2001) and Kim et al. (2005). This method
is effective in estimating the velocity dispersion of both
virialized systems and quasi-virialized systems (Xu et al. 2000).

Figure 9 plots the distribution of X-ray luminosity $L_{\rm x}$
versus velocity dispersion of dark matter $\sigma_{\rm dm}$ for
simulation sample C decomposed in cells on a scale of $1.04$ $h^{-1}$
Mpc. The observed $L_{\rm x}$ - $\sigma$ points of Mulchaey \& Zabludoff
(1998) are also shown in Figure 9. Many objects in this sample are
actually rich clusters, which are outside of the simulated data
points. Nevertheless, the observed points with low $L_{\rm x}$ and
$\sigma$ are clearly support to the simulated result.

The distribution of $L_{\rm x}$ versus $\sigma_{\rm dm}$ in Figure 9 is
also substantially scattered. However, it does dot have a very
clear upper envelopes but one can still see a less clear upper
envelop, which can be roughly fitted by a power law $L_{\rm
x}\propto \sigma^{2.1}$. That is, if we try to describe the
scattered $L_{\rm x}-\sigma$ distribution by a power law $L_{\rm
x}\propto \sigma^{\alpha_{LV}}$, the index $\alpha_{LV}$ should be
larger than 2.1. This is consistent with observational results
$\alpha_{LV}>2.3$ (Helsdon \& Ponman 2000; Xue \& Wu 2000). The
large scatter of $L_{\rm x}$ versus $\sigma$ is caused by the large scatter
of velocity versus density of both baryonic matter and dark matter.

\begin{deluxetable}{cccc}
\tablecolumns{8} \tablewidth{0pc} \tablecaption{Mean intensity of
soft-band X-ray background} \tablehead{ $\rho_{\rm dm}$&
\multicolumn{2}{c}{Mean intensity (ergs s$^{-1}$ cm$^{-2}$
deg$^{-2}$)}\cr\cline{2-3}($10^5<T<10^7$ K) & \multicolumn{1}
{c}{Primordial abundance} & \multicolumn{1}{c}{Metal abundance} }
\startdata
total & 1.73$\pm 0.63\times 10^{-12}$ & 1.87$\pm 0.63\times 10^{-12}$\\
$\rho_{\rm dm}<100$ & 2.90$\pm 0.62\times 10^{-13}$
    & 2.86$\pm 0.86\times
10^{-13}$\\
$\rho_{\rm dm}<50$ & 1.27$\pm 0.29 \times 10^{-13}$
     & 1.23$\pm 0.38\times 10^{-13}$\\
$\rho_{\rm dm}<10$ & 1.55$\pm 0.41\times 10^{-14}$
   & 1.52$\pm 0.63\times 10^{-14}$\\
\label{table:mx1}
\enddata
\end{deluxetable}

\section{Soft X-ray background}

\subsection{Mean and PDF of soft X-ray background }

As the hot gas with high temperature ($T>10^7$K) is always
identified as clusters, we calculate the X-ray background intensity
only from baryonic gas with temperature $10^5-10^7$K, which is given
by an integral
\begin{equation}
F(\nu_0)=
   \frac{1}{4\pi}\frac{c}{H_0}\int_{0}^{z_r}
   \epsilon([1+z]\nu_0, z)\frac{dz}
{(1+z)^4[\Omega_{\rm m}(1+z)^3+\Omega_{\Lambda}]^{1/2}},
\end{equation}
where $F(\nu_0)$ is in units of ergs s$^{-1}$ cm$^{-2}$ Hz$^{-1}$
sr$^{-1}$ and $\epsilon(\nu_0, z)$ is the volume emissivity of X-ray
photons with local frequency $\nu_0$ at redshift $z$. The emissivity
$\epsilon(\nu_0, z)$ is calculated using Raymond-Smith model 
(Raymond \& Smith 1977).
In the calculation of the X-ray background intensity, both the 
samples with and without metal cooling are considered.

The soft X-ray background mostly comes from hot clouds at $z<2$. It
is sufficient to produce maps by taking the integral of Eq.(7)
from $z=0$ to 6. The algorithm of mapping is as follows. The
simulation data are output every time when light crosses the
simulation box with a size of 100 h$^{-1}$ Mpc. For each output,  we
take the integral in Eq.(7) along one random chosen axis and produce
a map of X-ray flux $F$ at the redshift $z$ of that box. After the
simulation reaches $z=0$, we have 60 maps of $F$ at redshifts
corresponding to a comoving distance at $n\times 100$ $h^{-1}$ Mpc,
where $n=0, ... 59$. One map of the background can then be obtained by
(1) arranging a set of 60 two-dimensional maps with a randomized center, 
(2) taking a constant angular projection for each map, and (3)
superposing the contributions of the 60 maps.

A typical two-dimensional map of the soft X-ray background from regions
with $10^5<T<10^7$ K based on sample A is shown in Figure 10. The
angular size of the map is $1^{\circ}\times 1^{\circ}$, and its
angular resolution is $0'.06$. Figure 11 gives the corresponding
PDF of the intensity of the X-ray background in the band $0.5 - 2$
keV. For sample A, the mean X-ray background intensity is
$(1.87\pm 0.63)\times 10^{-12}$ ergs s$^{-1}$ cm$^{-2}$ deg$^{-2}$,
where the 1 $\sigma$ error is from the variance of 50 two-dimensional maps.
For sample B, we have $(1.73 \pm 0.63)\times 10^{-12}$ ergs
s$^{-1}$. Therefore, the effect of metal cooling has to be less
than 10\%.

\subsection{Density dependence of soft-band X-ray background}

It is usually believed that most X-ray background is attributed to
resolvable discrete X-ray sources, such as active galactic nuclei (AGNs) 
and truly diffuse
extragalactic baryonic clouds. As shown in \S 3, the
dynamical and thermodynamical properties of baryonic clouds are
continuously varying from low-density (diffuse) regions to
high-density regions (collapsed and virialized objects). There is
no clear criterion to distinguish the X-ray emission of truly
diffuse baryonic gas from that of resolvable sources. We 
study this problem with the density dependence of the soft X-ray
background.

We integrate Eq.(7), but only calculate X-ray emission from
baryonic clouds with temperatures $10^5<T<10^7$K and in regions
with $\rho_{\rm dm}<$100, 50, and 10. The results for
sample A are shown in Figure 12, which contains three
$1^{\circ}\times 1^{\circ}$ maps corresponding to $\rho_{\rm
dm}<$100, 50, and 10. The mean intensities of the 
soft-band X-ray
background are listed in Table 2. We can see that less than 17\%
(7\%) of the total soft X-ray background of $10^5<T<10^7$ comes
from clouds located in regions of dark matter with density
contrast $\rho_{\rm dm}<100$ (50). As mentioned in \S 3.2, the
cells having $\rho_{\rm dm}>100$ or $>50$ generally contain
collapsed and virialized structures. From Figure 2 we can also see
that the structures with $\rho_{\rm dm}>100$ or 50 have only a very
small volume fraction. These regions are discretely distributed.
Therefore, they should be resolvable. Thus, one can conclude that
the contribution of {\it truly} diffuse baryonic gas to the
soft-band X-ray background is no more than about $2.90\times
10^{-13}$ ergs s$^{-1}$ cm$^{-2}$ deg$^{-2}$. This result can also
be seen from Table 1. Using data of the $XMM-Newton$ Lockman Hole
observation, Worsley et al (2005) show that in the soft-band ($<$2
keV), more than 90$\%$ of the X-ray background can be
resolved. This result is consistent with Table 2. The simulation
done by Croft et al (2001) found the X-ray background of 
the warm-hot intergalactic medium (WHIM) to
be $4.15\times 10^{13}$ ergs s$^{-1}$ cm$^{-2}$ deg$^{-2}$. This is a
little larger than our result. This difference is probably
attributable to the powerful shock-capturing ability of the WENO
code.

It is also interesting to note from Table 2 that for the case of
the total region, the sample considering the metal abundance (sample A) 
has a slightly
higher mean intensity of the soft X-ray background than that without
metal cooling (sample B), while for all other cases $(\rho_{\rm
dm}<100)$ sample A gives a slightly lower mean intensity than
that of the primordial abundance. This is because metal cooling can
weaken the turbulence heating. The gas with metal cooling easily falls 
into gravity wells ($\rho_{\rm dm}>100)$. On the
other hand, in the case of no metal cooling more gas will remain
in less clustered regions and give higher X-ray emission from
clouds with $\rho_{\rm dm}<100$. In all cases, the difference between
samples A and B is small. Thus, the thermodynamical effect of metal
cooling is much less the turbulence heating.

\subsection{Temperature dependence of soft X-ray background}

Similarly, we study the temperature dependence of soft X-ray
background. We do the integral in Eq.(7), but only calculate X-ray
emissions from baryonic clouds with temperatures less than
$10^{6.5}$ K. A $1^{\circ}\times 1^{\circ}$ map of the soft X-ray
background from clouds of $T<10^{6.5}$ of sample A is shown in
Figure 13, and the corresponding PDF of the intensity of the X-ray
background is given in Figure 14. The mean intensity is $(9.38\pm
3.06)\times 10^{-14}$ ergs s$^{-1}$ cm$^{-2}$ deg$^{-2}$ for sample
A and $(8.0\pm 2.03)\times 10^{-14}$ ergs s$^{-1}$ cm$^{-2}$
deg$^{-2}$ for sample B. Therefore, no more than 5\% of the total
soft X-ray background of $10^5<T<10^7$K comes from clouds with
temperature $T < 10^{6.5}$ K. Most of soft X-ray background is
attributed to baryonic clouds with high temperature
($T>10^{6.5}$).

\section{Discussion and conclusions}

The evolution of baryonic gas can be roughly divided into three
stages. The first is the linear stage. It can be simply described
by the similarity between the baryonic and dark matter. When the
non-linear evolution takes place, Burgers turbulence develops.
Baryonic gas will decouple with dark matter, and becomes
multiphase. The kinetic energy of baryon fluid will be
dissipated by the shocks of turbulence. Some cosmic baryonic gas
finally falls into massive halos and evolves in the thermal
equilibrium state in gravity wells. Some heated baryonic gas
remains outside of gravity wells. The Burgers turbulence heating
is not uniform. A uniform heating would contradict with the Ly$\alpha$
forest, which is produced by hydrogen clouds with a temperature of
about $10^4$-$10^5$K. Therefore, baryonic gas undergoing the Burgers
turbulence evolution must be multiphase. In the stage of the
Burgers turbulence, the dominant component in volume fraction is
the gas in the low-temperature phase (He et al. 2004).

We show that the X-ray emission of baryonic gas can be well
modeled by the above-mentioned scenario. That is, the observed
distributions of $L_{\rm x}$ versus $T$ and $L_{\rm x}$ versus 
$\sigma$ of galaxy
groups can be reproduced by the simulation sample of the $\Lambda$CDM
model without adding extra heating sources. We also find that
almost all of the soft-band X-ray background radiation is from
clustered regions, and the contribution of diffuse gas with $\rho_{\rm
dm}<50$ is negligible. This is also consistent with current X-ray
observations. These results show that the Burgers turbulence
heating alone seems to be enough to account for the basic features
of X-ray emission of baryonic gas in the universe.

We should also point out that star formation and its feedback on
the baryonic gas are not considered in our simulation. Roughly,
there are two types of feedback: (1) photoionization heating by
the UV emission of stars and AGNs and (2) injection of hot gas and
energy by supernova explosions, or other sources of cosmic rays.
Actually, the photoionization heating can be properly considered,
if the UV background is adjusted by fitting the simulation with
the observed mean flux decrement of QSOs' Ly$\alpha$ absorption
spectrum (Feng et al. 2003). The significant effect of injecting
hot gas and energy by supernovae is mostly on dwarf galaxies.
Recent Ly$\alpha$ observations of protoclusters (Adelberger et al.
2003) suggest that AGN heating does not drastically affect the gas
in clusters. Therefore, this heating mechanism may not be strong
enough to change the basic features of the multiphase scenario
given in this paper. Moreover, the shocks may lead to the electron
temperature $T_{\rm e}$ being lower than the ion temperature $T_{\rm i}$.
Since the mean of the ratio $T_{\rm i}/T_{\rm e}$ is generally less than 3
(Fox \& Loeb 1997; Takizawa 1998; Yoshida et al. 2005), this
effect may lead to an uncertainty of the luminosity $L_{\rm x}$ that is 
less than a factor of 2. Therefore, all conclusions should basically
be held.

\acknowledgments
We thank the anonymous referees for many valuable
suggestions. T.-J.Z. is supported by the Fellowship of the World
Laboratory and thanks Peng-Jie Zhang for useful discussion and
Ue-Li Pen for help. J.-R.L. is appreciative of Hy Trac for giving
the parallel gravitation solver code and Cheng Ling-Mei for help on
the Raymond-Smith model. T.-J.Z., L.-L.F. and P.H. acknowledge support from
the National Science Foundation of China (grants 10473002,
10533010, 10573036 and 10545002). This work was also partially
supported by US NSF AST 05-07340.

\newpage

\begin{figure}
\centering
\includegraphics[width=12.1cm,angle=270]{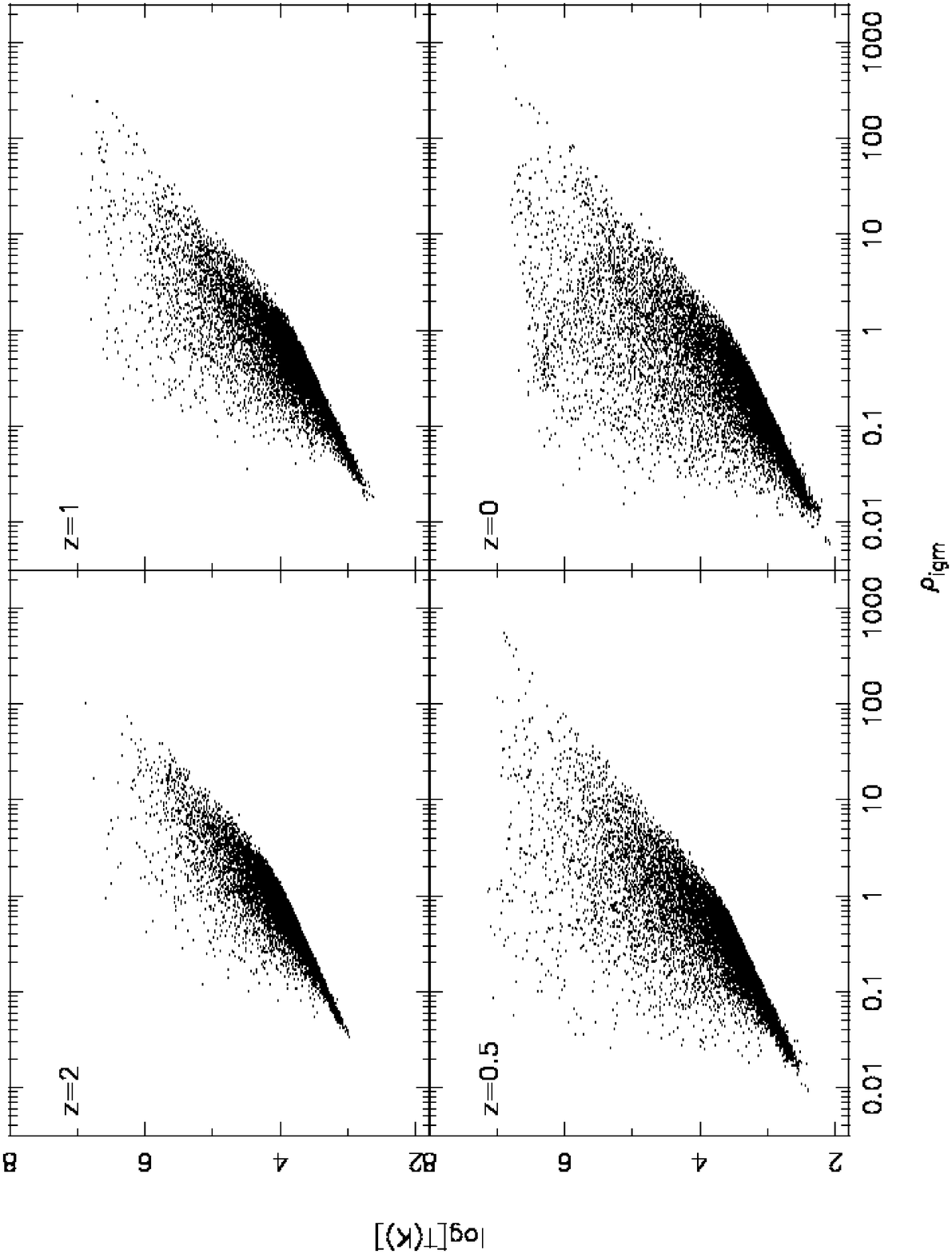}
\caption {Temperature $T$ vs. baryon density
$\rho_{\rm igm}$ for sample C at redshifts 2, 1, 0.5, and 0, where
$\rho_{\rm igm}$ is in units of the mean baryonic matter density
$\bar\rho_{\rm igm}$. Each panel contains 19,200 data points randomly
selected from sample C.}
\end{figure}

\newpage

\begin{figure}
\centering
\includegraphics[width=8.0cm,angle=270]{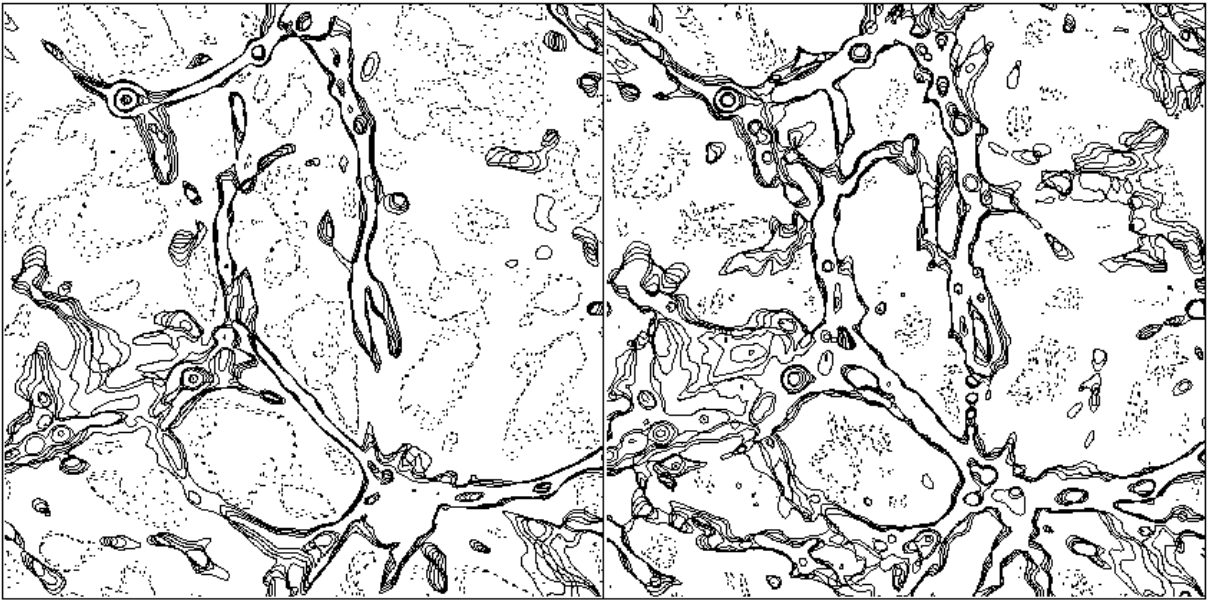}
\caption {Density contour plots of dark matter ({\it left}) and
            baryonic gas ({\it right}) for a slice of 0.26 $h^{-1}$
            Mpc thickness at $z=0$. The solid contours encompass
            the over-dense regions with $\rho=e^{i/2}$, $i=0,1,2...$
            ($\bar{\rho}$ is normalized to 1), while the dotted lines
            represent the boundaries of the under-dense regions with
            $\rho=e^{-i/2}$, $i=1,2...$.}
\end{figure}

\newpage

\begin{figure}
\centering
\includegraphics[width=8.0cm,angle=270]{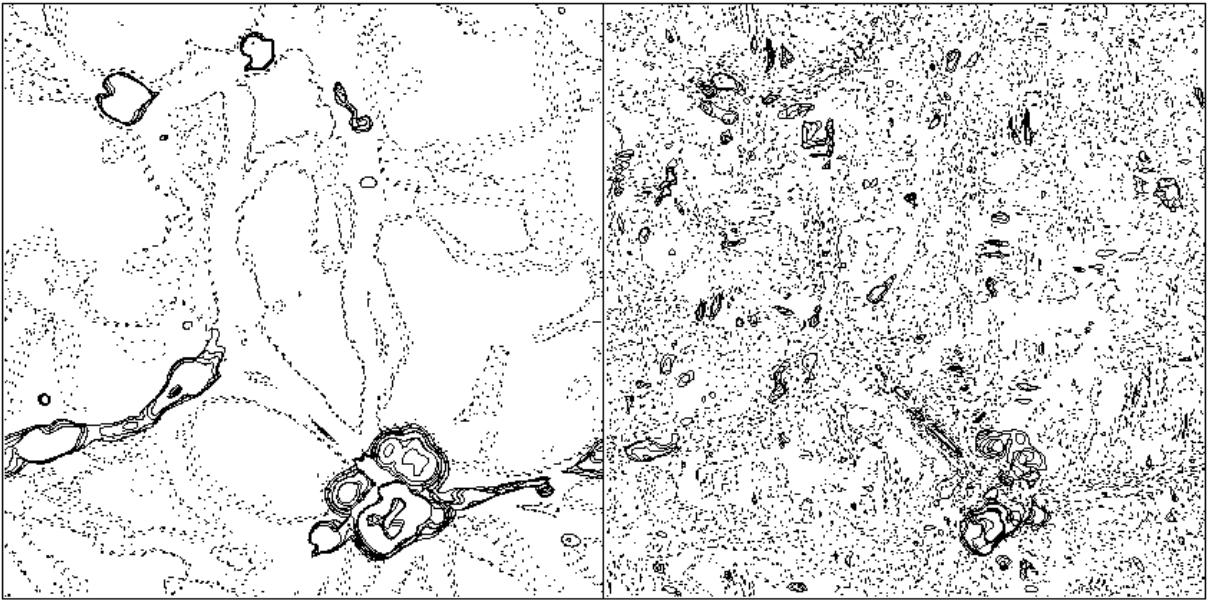}
\caption {Contour plots of baryon to dark ratio
$\rho_{\rm igm}/\rho_{\rm dm}$
({\it left}) and temperature $T$ ({\it right}) for the same slice
as in Figure 2.
The solid contours represent, respectively,
the regions with $\rho_{\rm igm}/\rho_{\rm dm}=e^{i/3}, i=0,1,2,...$,
and $T=e^{i/2}\times 10^5$ K, $i=0,1,2,...$, while the dotted lines
represent $\rho_{\rm igm}/\rho_{\rm dm}=e^{-i/3}$, $i=1,2,...$, and
$T=e^{-i/2}\times10^5$ K, $i=1,2,...$, regions.}
\end{figure}

\newpage

\begin{figure}
\centering
\includegraphics[width=12.1cm,angle=270]{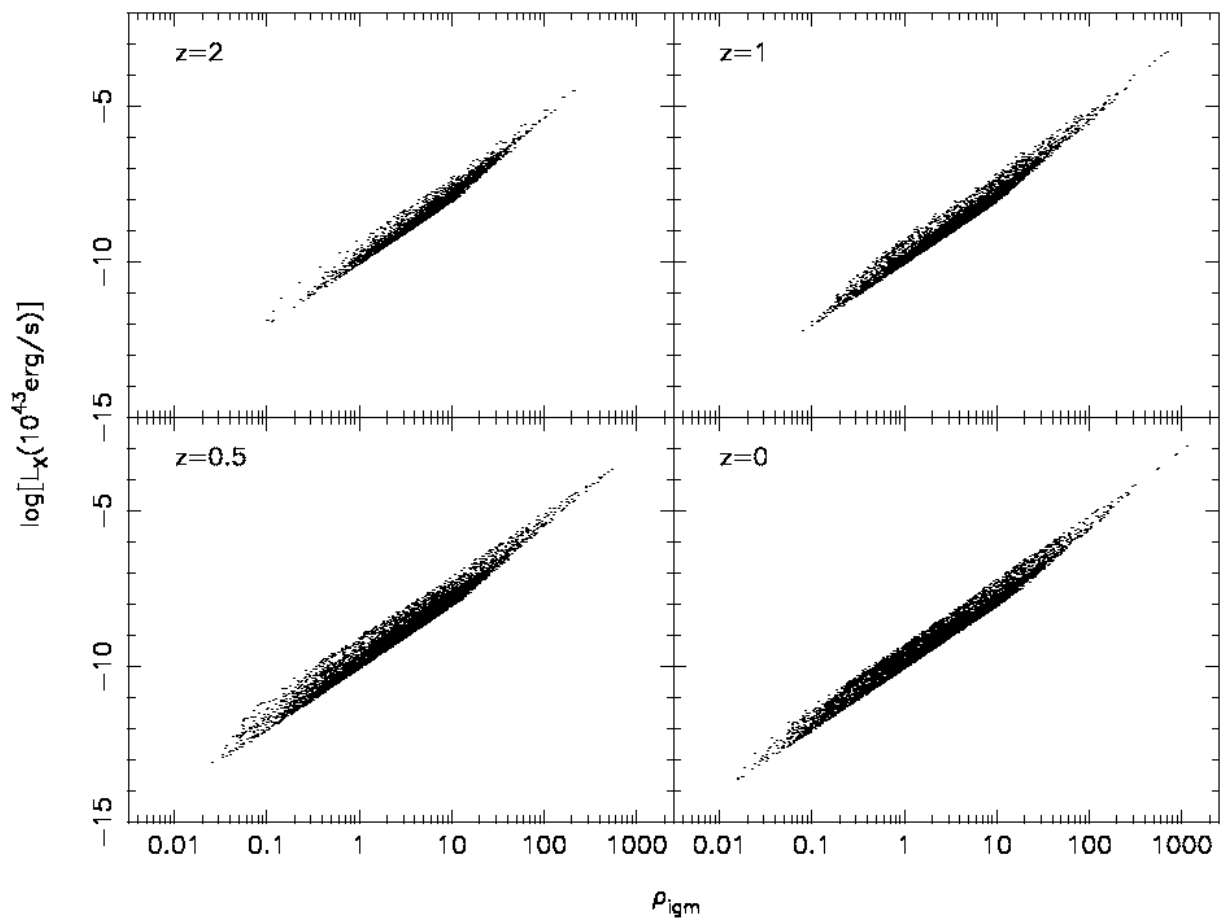}
\caption {X-ray luminosity $L_{\rm x}$ vs. baryon density
$\rho_{\rm igm}$ for sample C, where $\rho_{\rm igm}$ is in units of
the mean baryonic matter density $\bar\rho_{\rm igm}$. The panels at
$z$=0.0, 0.5, 1.0, and 2 show, respectively, 9931, 9119, 7314, and
4864 data points randomly selected from sample C in each redshift,
and with temperature $T>10^5$ K.}
\end{figure}

\newpage

\begin{figure}
\centering
\includegraphics[width=19.cm,angle=270]{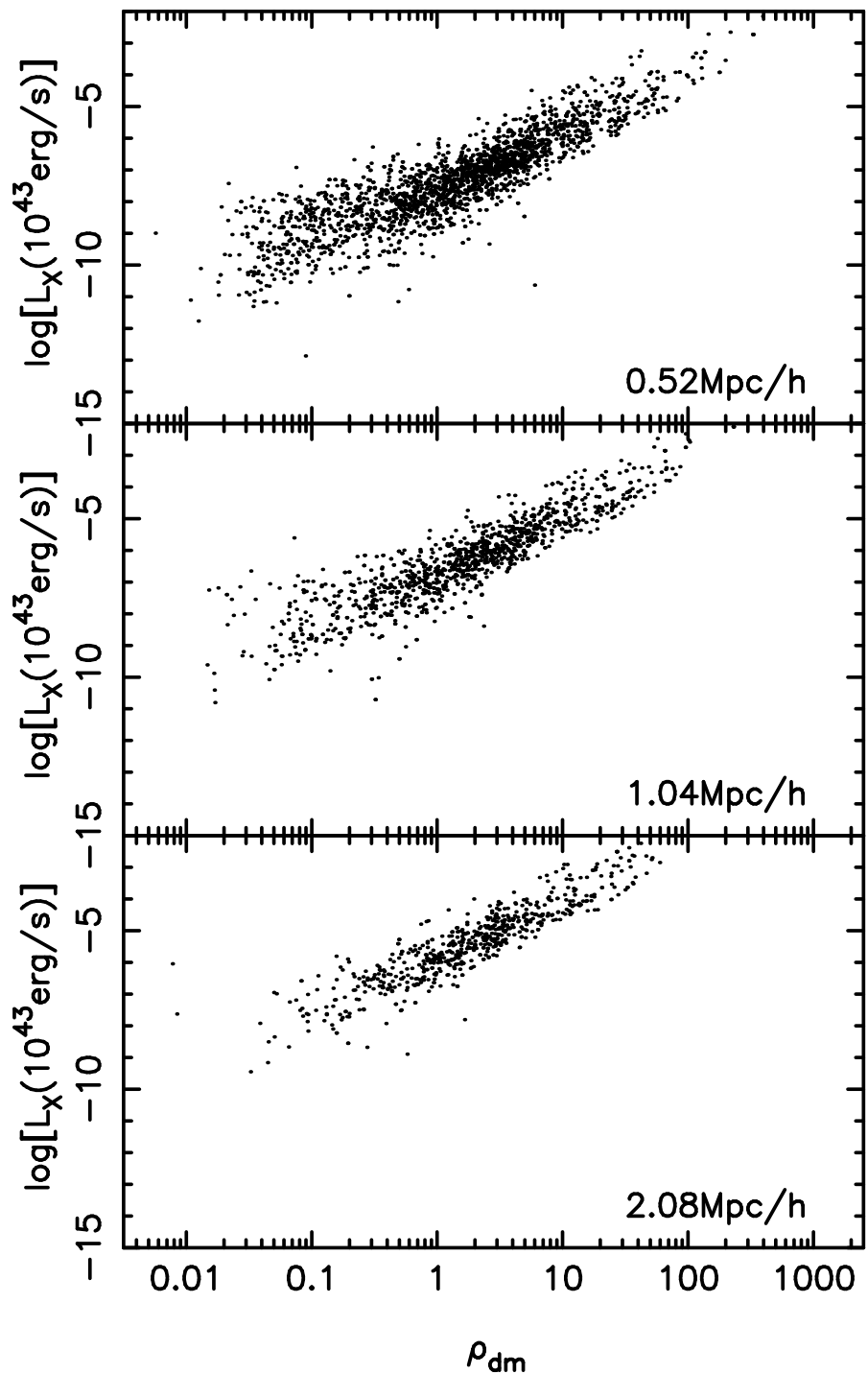}
\caption {X-ray luminosity $L_{\rm x}$ vs. dark matter density
$\rho_{\rm dm}$ for sample C at redshift $z=0$ and
decomposed on scales of 0.52, 1.04, and 2.08 $h^{-1}$ Mpc. 
Here $\rho_{\rm dm}$ is in units of the mean dark matter 
density $\bar\rho_{\rm dm}$.}
\end{figure}

\newpage
\begin{figure}
\centering
\includegraphics[width=11.5cm,angle=270]{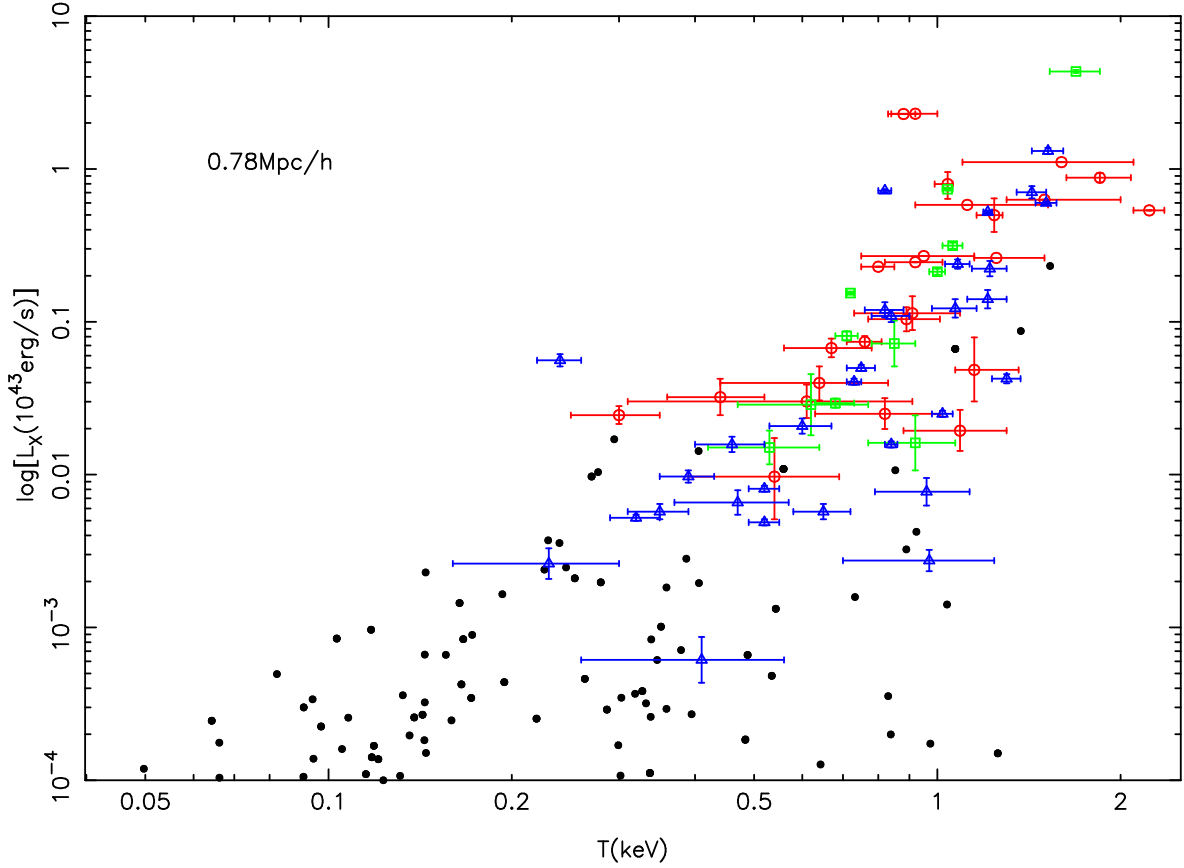}
\caption {X-ray luminosity $L_{\rm x}$ vs. temperature $T$ for
samples A at redshift $z=0$ with the DWT decomposition on scales
of 0.78 $h^{-1}$ Mpc. The observed data are adapted from Helsdson \&
Ponman (2000; {\it squares}), Xue \& Wu (2000; {\it circles}) and Croston et al
(2005; {\it triangles}).}
\end{figure}

\newpage

\begin{figure}
\centering
\includegraphics[angle=270]{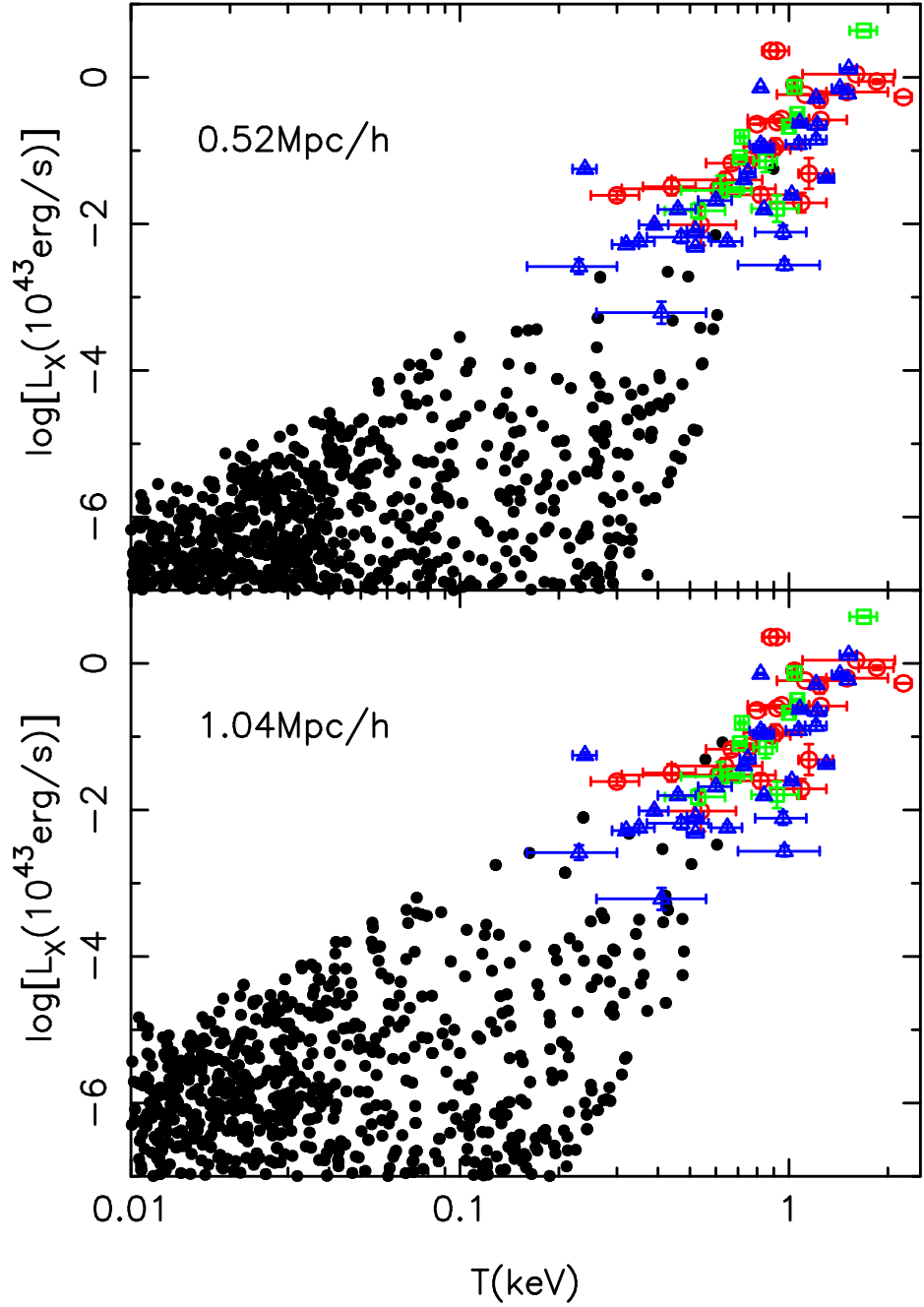}
\caption {Same as Fig.6, but for sample C at redshift $z=0$ with
the DWT decomposition on scales of 0.52 ({\it top}) and 1.04 $h^{-1}$ 
Mpc ({\it bottom}).}
\end{figure}

\newpage

\begin{figure}
\centering
\includegraphics[angle=270]{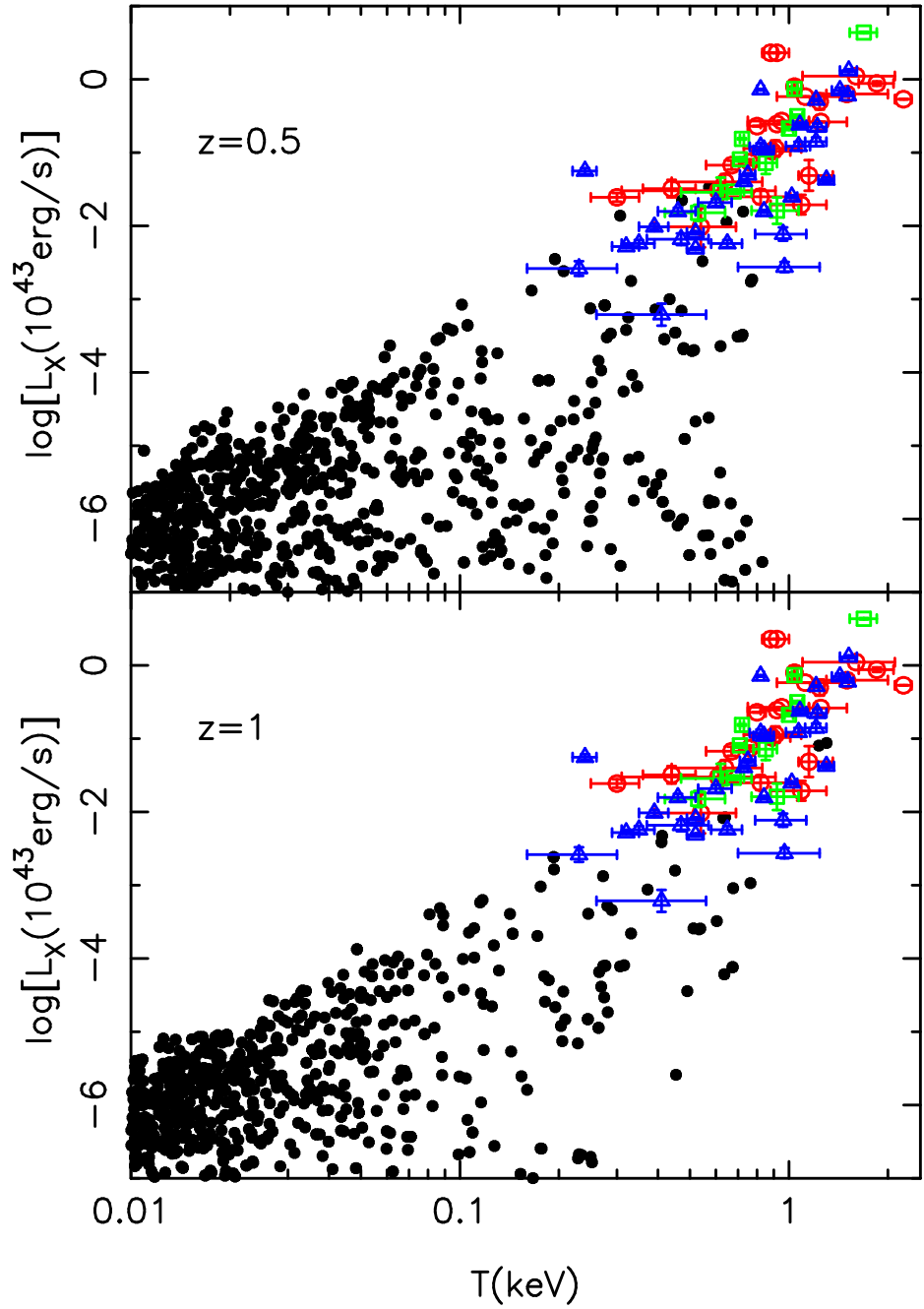}
\caption {Same as Fig.7, but for redshift $z=0.5$ ({\it top}) and
$z=1$ ({\it bottom}). The physical scale of the DWT decomposition is
0.69 $h^{-1}$ Mpc. }
\end{figure}

\newpage
\begin{figure}
\centering
\includegraphics[width=11.5cm,angle=270]{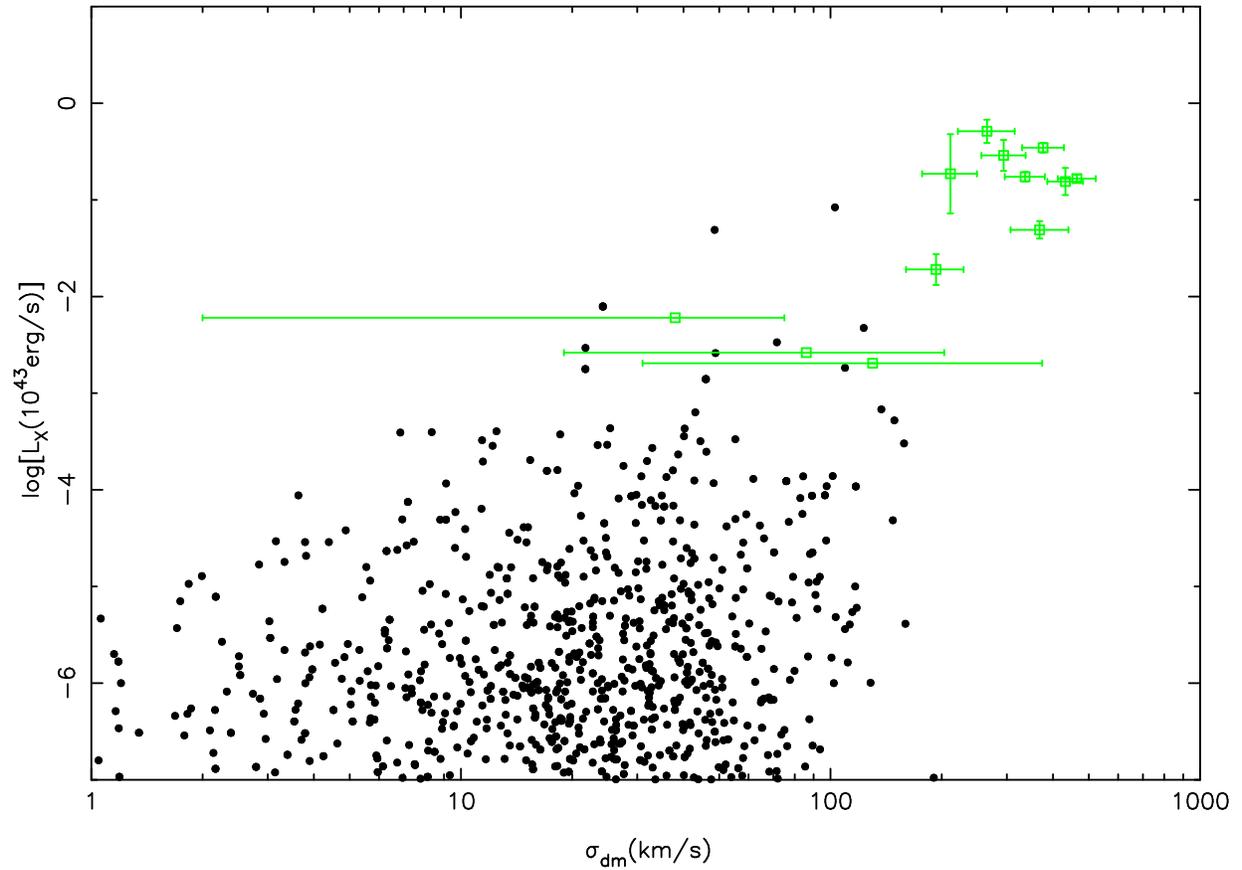}
\caption {X-ray luminosity $L_{\rm x}$ vs. velocity dispersion of
dark matter $\sigma_{\rm dm}$ for sample C at redshift $z=0$ with
the DWT decomposition on scales of 1.04 $h^{-1}$ Mpc. The observed data are
adapted from Mulchaey \& Zabludoff(1998).}
\end{figure}

\newpage

\begin{figure}
\centering \epsscale{1.} \plotone{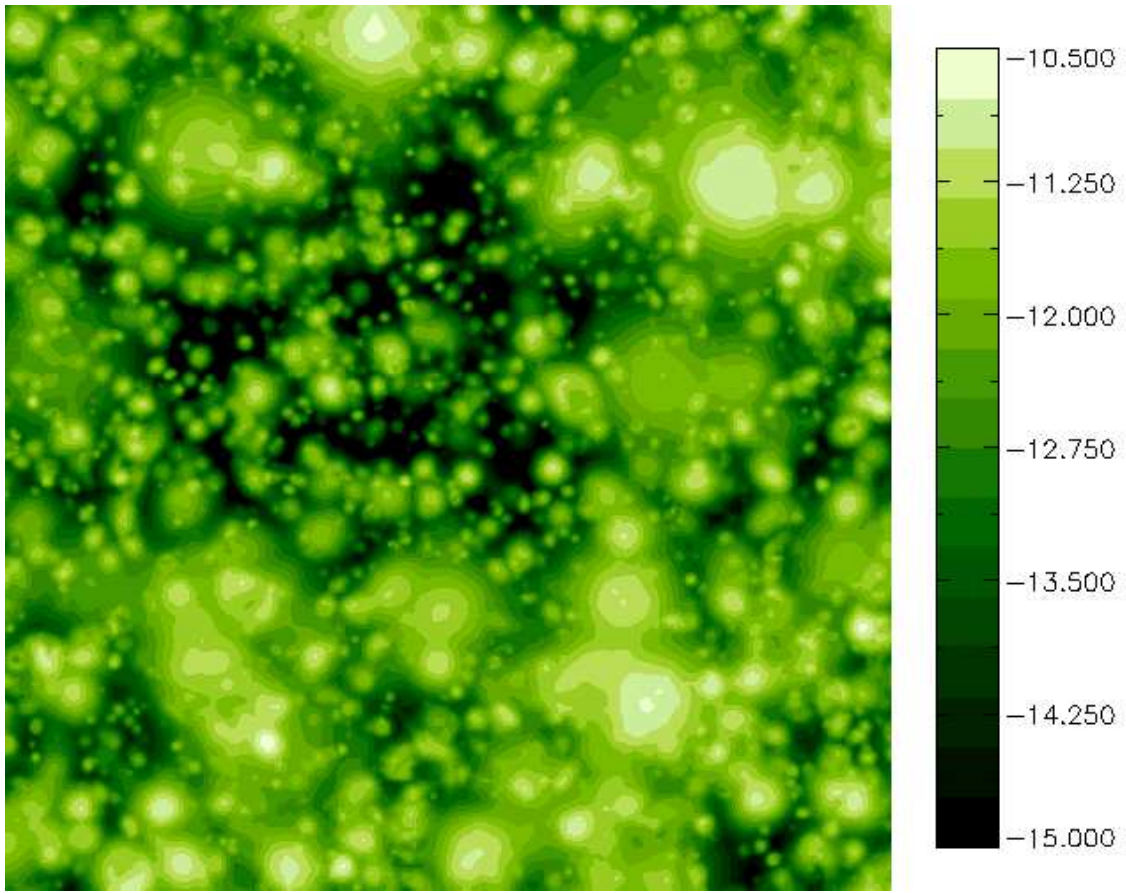}
\caption{A $1^{\circ}\times 1^{\circ}$ map of X-ray intensity in
soft 0.5-2 keV band for simulation sample A.}
\end{figure}

\newpage
\begin{figure}
\centering
\includegraphics[width=11.5cm,angle=270]{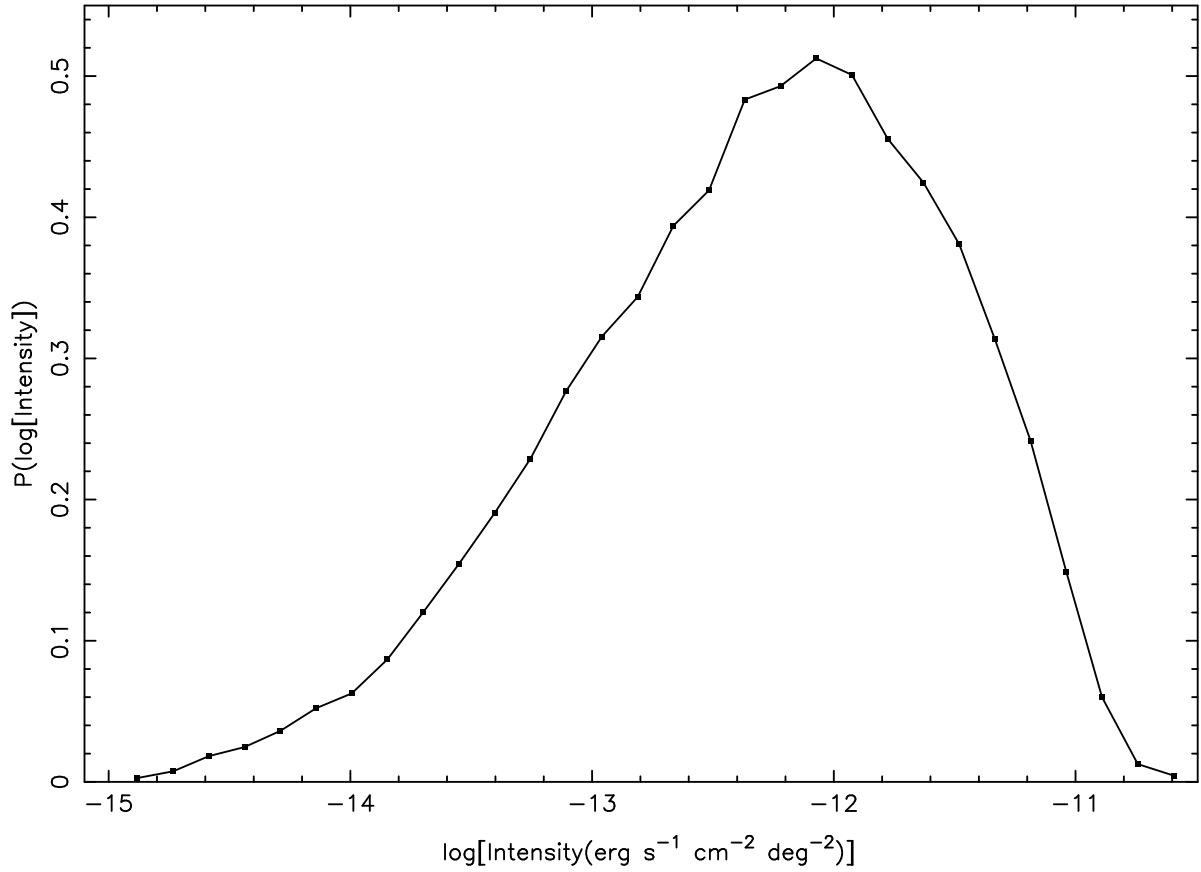}
\caption {Probability distribution function of X-ray intensity in
the 0.5-2 keV band for the same samples as in Fig.10.}
\end{figure}

\newpage

\begin{figure}
\centering
\includegraphics[width=8.5cm,angle=0]{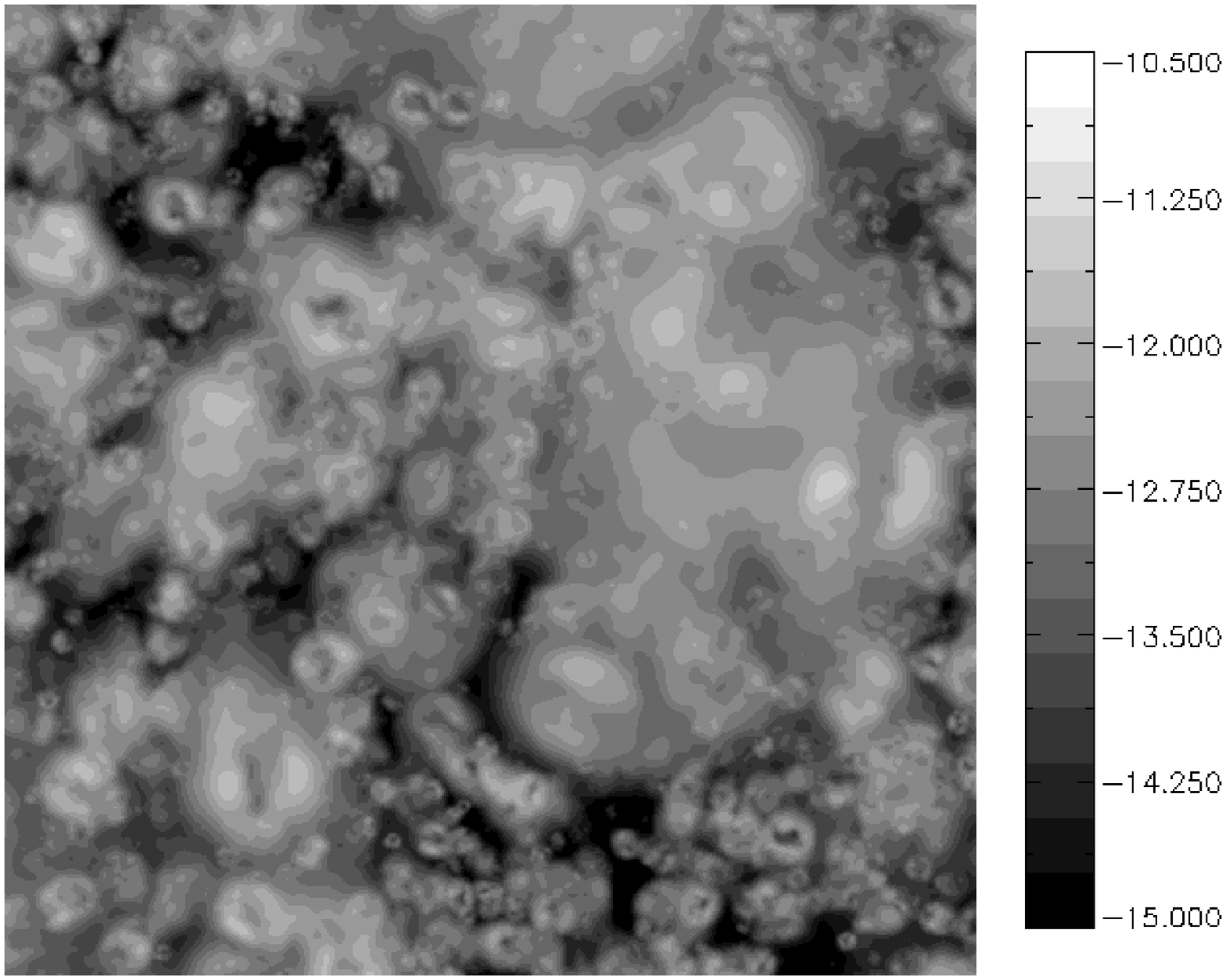}
\includegraphics[width=8.5cm,angle=0]{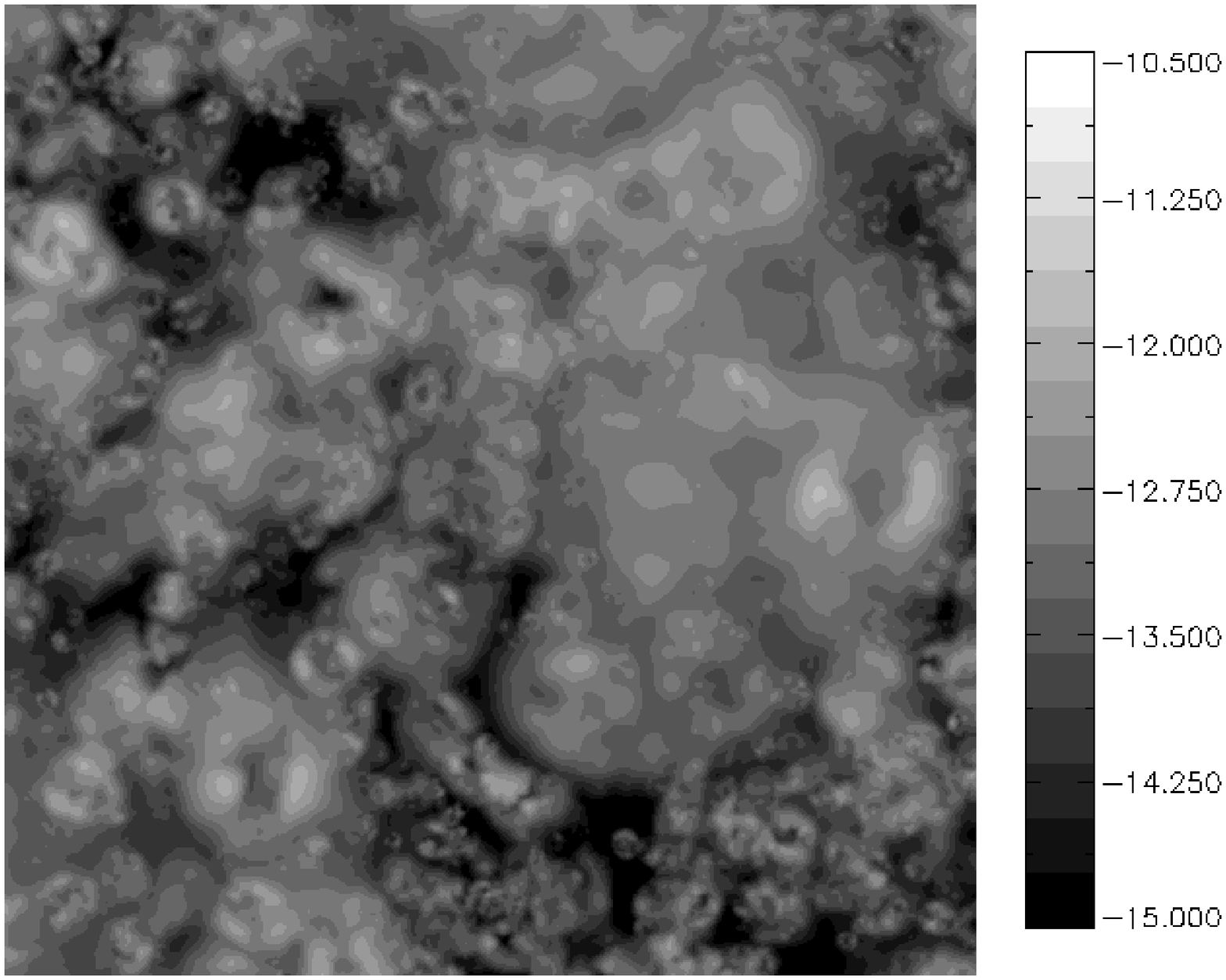}
\includegraphics[width=8.5cm,angle=0]{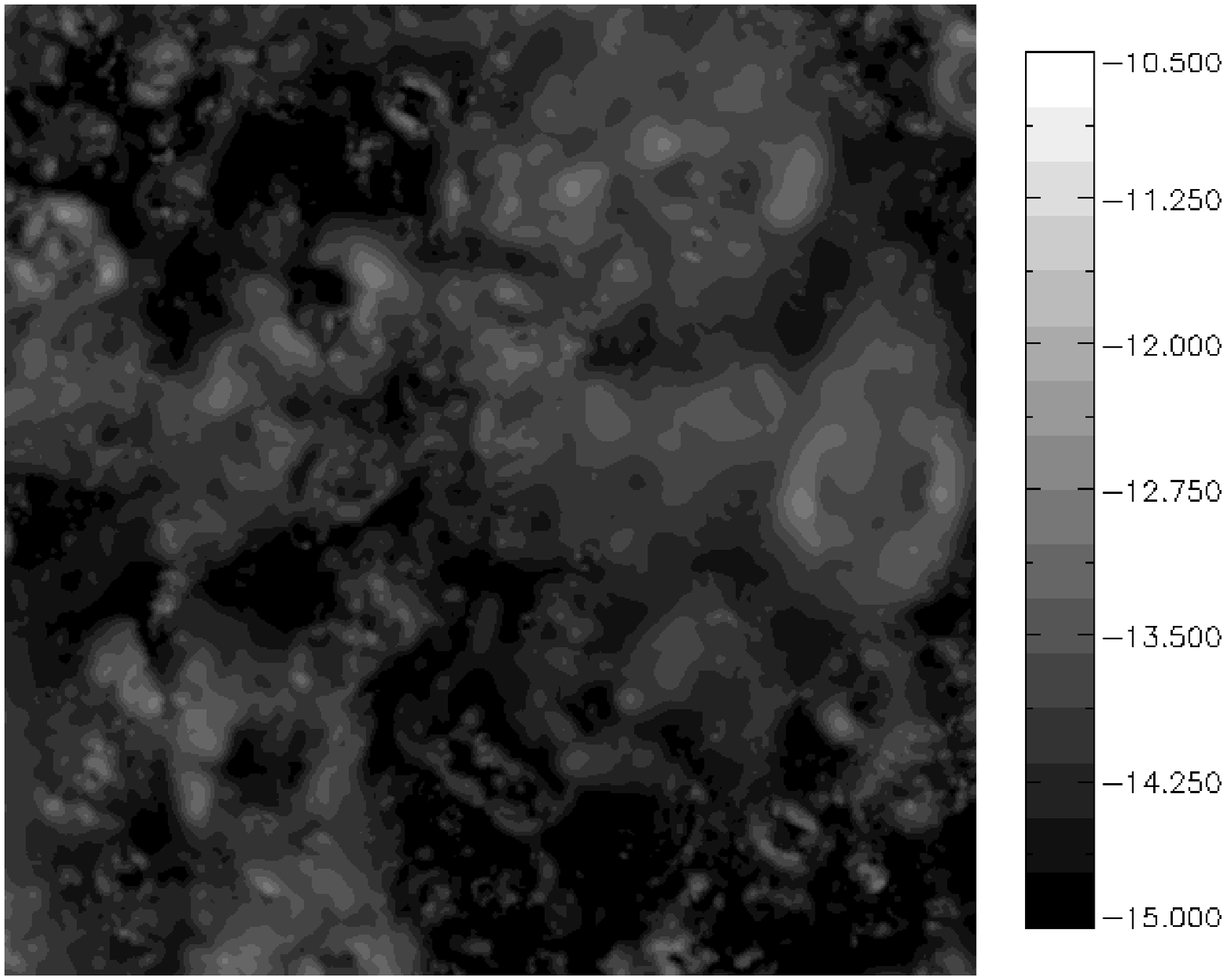}
\caption{Same as Fig.10, but for baryonic clouds located in
regions with dark matter density $\rho_{\rm dm}<100$ ({\it top}), 50 
({\it middle}) and 10 ({\it bottom}), where $\rho_{\rm dm}$ is in units 
of the mean dark matter density $\bar\rho_{\rm dm}$.}
\end{figure}

\newpage

\begin{figure}
\centering 
\epsscale{1.} 
\plotone{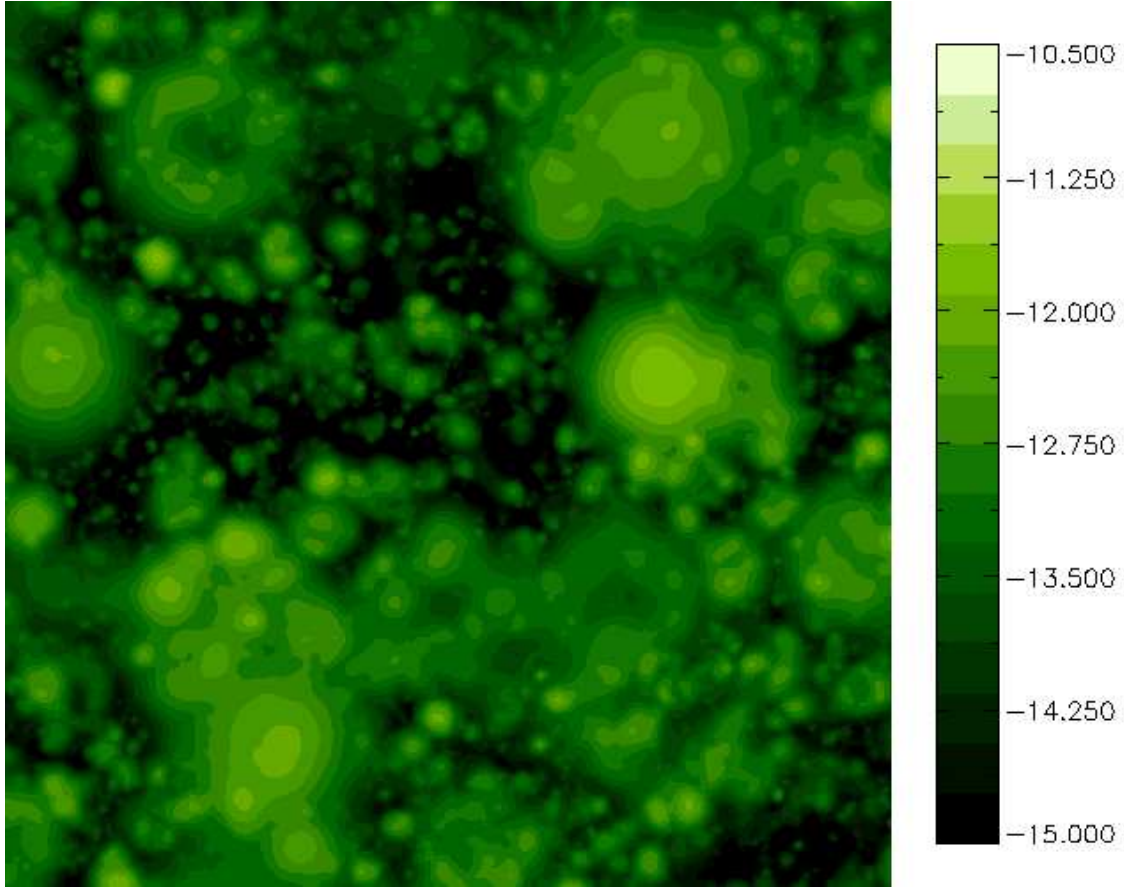} 
\caption{Same as Fig.10, but for baryonic clouds with temperature in
the range $10^5<T<10^{6.5}$K.}
\end{figure}

\newpage

\begin{figure}
\centering
\includegraphics[width=11.5cm,angle=270]{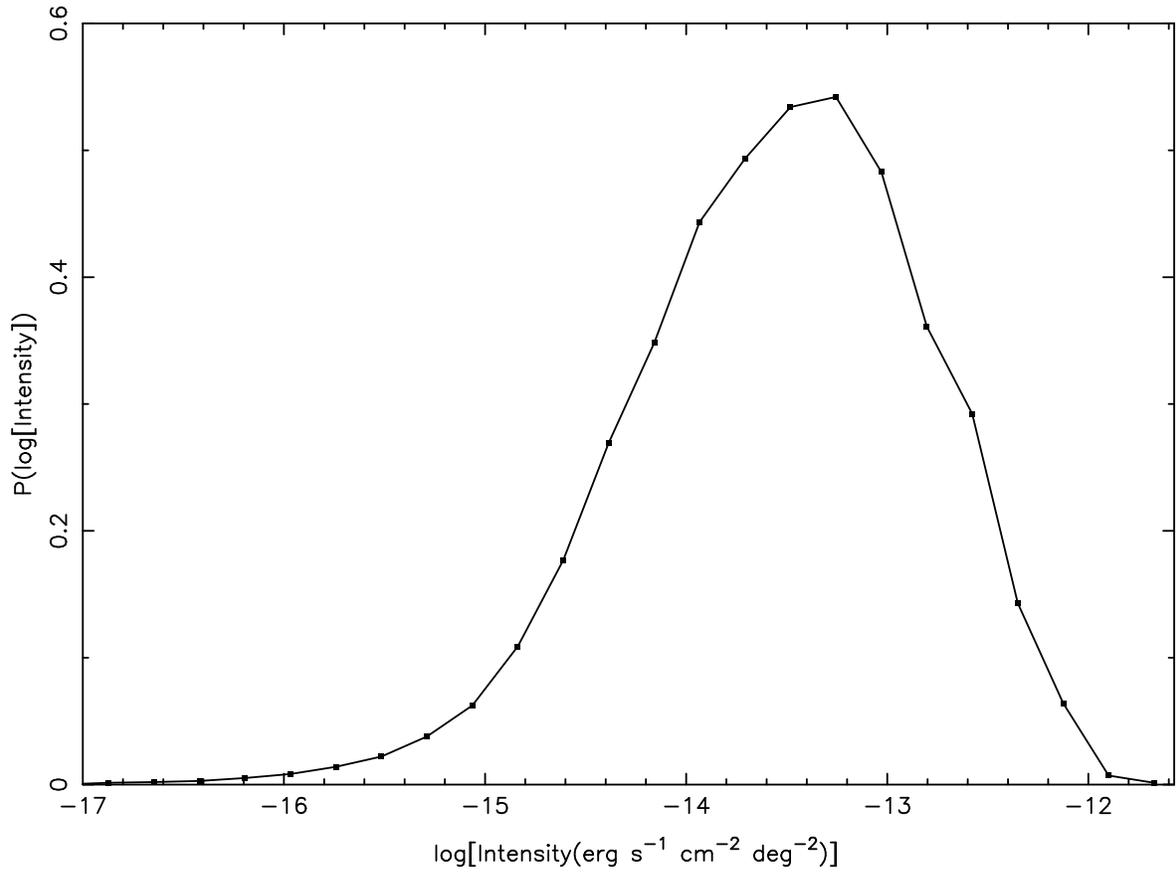}
\caption{Same as Fig.11, but for the x-ray emission from 
clouds with temperature in range $10^5<T< 10^{6.5}$K (the same samples as in Fig.13).}
\end{figure}

\end{document}